\documentclass[aps,prd,floatfix,superscriptaddress,showpacs,showkeys]{revtex4}

\usepackage{amssymb,amsmath,amsfonts,latexsym,graphicx,epsfig}

\RequirePackage{xspace}

\usepackage{titlesec}
\usepackage{color}
\titleformat{\section}{\large\bfseries}{\thesection}{1em}{}

\newcommand{\bea}{\begin{eqnarray}}
\newcommand{\ena}{\end{eqnarray}}
\newcommand{\be}{\begin{equation}}
\newcommand{\en}{\end{equation}}

\begin{document}

\title{On the strong effective coupling, glueball and meson ground states}

\author{Gurjav Ganbold}
\affiliation{Bogoliubov Laboratory of Theoretical Physics, JINR, \\
                         Joliot-Curie 6, 141980 Dubna, Russia}
\affiliation{Institute of Physics and Technology, Mongolian Academy of Sciences, \\
Enkh Taivan 54b, 13330 Ulaanbaatar, Mongolia}

\begin{abstract}
The phenomena of strong running coupling and hadron mass generating have
been studied in the framework of a QCD-inspired relativistic model of quark-gluon
interaction with infrared confined propagators.
We derived a meson mass equation and revealed a specific new behaviour of the
mass-dependent strong coupling $\hat\alpha_s(M)$ defined in the time-like region.
A new infrared freezing point $\hat\alpha_s(0)=1.03198$ at origin has been found
and it did not depend on the confinement scale $\Lambda>0$. Independent and
new estimates on the scalar glueball mass, 'radius' and gluon condensate value
have been performed. The spectrum of conventional mesons have been
calculated by introducing a minimal set of parameters: the masses of constituent
quarks and $\Lambda$. The obtained values are in good agreement with the latest
experimental data with relative errors less than 1.8 percent. Accurate estimates
of the leptonic decay constants of pseudoscalar and vector mesons have been
performed.
\end{abstract}


\maketitle

\section{INTRODUCTION}

The low-energy region below $1\div 2$ GeV becomes a testing ground, where
many novel, interesting and challenging behavior is revealed in particle physics
(see, e.g., \cite{PDG16}). Any QCD-inspired theoretical model should be able to
describe correctly hadron phenomena such as confinement, running coupling,
hadronization,  mass generation etc. at large distances. The inefficiency of the
conventional perturbation theory in low-energy domain pushes particle physicists
to develop and use different phenomenological and nonperturbative approaches,
such as QCD sum rule \cite{shif79}, chiral perturbation theory \cite{gass84}, heavy
quark effective theory \cite{neub94}, rigorous lattice QCD simulations \cite{gupt98},
the coupled Schwinger-Dyson equation \cite{robe03} etc.

The confinement conception explaining the non-observation of color charged
particles (quarks, gluons) is a crucial feature of QCD and a great number of
theoretical models have been suggested to explain the origin of confinement.
Particularly, the confinement may be parameterized by introducing entire-analytic
and asymptotically free propagators \cite{leut81,stin84}), vacuum gluon fields
serving as the true minimum of the QCD effective potential \cite{elis85},
self-dual vacuum gluon fields leading to the confined propagators \cite{efim95},
the Wilson loop techniques \cite{kraa98}, lattice Monte-Carlo simulations
\cite{lenz04},  a string theory quantized in higher  dimensions \cite{alko07} etc.
Each approach has its benefits, justifications and limitations. A simple and
reliable working tool implementing the confinement concept is still required.

The strength of quark-gluon interaction $g$ in QCD depends on the mass scale
or momentum transfer $Q$. This dependence is described theoretically by the
renormalization group equations and the behavior of $\alpha_s\doteq g^2/(4\pi)$
at short distances (for high $Q^2$), where asymptotic freedom appears, is well
investigated \cite{beth00,chek03,pros07,beth09,PDG16,deur16} and measured,
e.g.,  $\alpha_s(M_Z^2)=0.1185 \pm 0.0006$ at mass scale $M_Z=91.19$ GeV
\cite{PDG14}. On the other hand, it is necessary to know the long-distance
(for $Q^2\le 1$ GeV) or, infrared (IR),behavior of $\alpha_s$ in order to
understand quark confinement, hadronization processes and hadronic structure.
Many phenomena in particle physics are affected by the long-distance property
of the strong coupling \cite{shir02,nest03,cour11,crew15,deur16},  however the
IR behavior of $\alpha_s$ has not been well defined yet, it needs to be more
specified. A self-consistent and physically meaningful prediction of  $\alpha_s$
in the IR region is necessary.

The existence of extra isoscalar mesons is predicted by QCD and in case of the
pure gauge theory they contain only gluons, and are called the {\sl glueballs},
the bound states of gluons. Nowadays, glueballs are the most unusual particles
predicted by theory, but not found experimentally yet \cite{klem07,PDG16}. The
study of glueballs currently is performed either within effective models or lattice
QCD. The glueball spectrum has  been studied by using the QCD sum rules
\cite{nari00},  Coulomb gauge QCD  \cite{szcz03},  variuos potential models
\cite{corn83,kaid06,math06}. Different string models are used for describing
glueballs \cite{solo01}, including  combinations of string and potential approaches
\cite{brau04}.  A proper inclusion of the  helicity degrees of freedom can improve
the compatibility between lattice QCD and potential models  \cite{math06a}.
Recent lattice calculations, QCD sum rules, 'tube' and  constituent 'glue' models
predict that the lightest glueball takes the quantum numbers ($J^{PC}=0^{++}$)
\cite{morn99,meye05,ochs06}. However, errors on the mass predictions are
large, particularly,  $M_G=1750\pm 50 \pm 80$  MeV for the mass of scalar
glueball from quenched QCD \cite{morn99}. Therefore, an accurate prediction
of the glueball mass combined with other reasonable unquenched estimates
and performed within a theoretical model with fixed global parameters is important.

One of the puzzles of hadron physics is the origin of the hadron masses. The
Standard Model and, in particular, QCD operate only with fundamental particles
(quarks, leptons, neutrinos), gauge bosons and the Higgs. It is not yet clear how
to explain the appearance of the multitude of observed hadrons and elucidate
the generation of their masses. Physicists have proposed a number of models
that advocate different mechanism of the origin of mass from the most
fundamental laws of physics.  The calculation of the hadron mass spectrum
in a quality comparable to the precision of experimental data remains actual.

In some cases, it is useful to investigate the corresponding low-energy effective
theories instead of tackling the fundamental theory itself. Indeed, data
interpretations and calculations of hadron characteristics are frequently carried
out with the help of phenomenological models.

One of the phenomenological approaches is the model of induced quark currents.
It is based on the hypothesis that the QCD vacuum is realized by the anti-selfdual
homogeneous gluon field~\cite{burd96}. The confining properties of the vacuum
field and  chiral symmetry breaking can explain the distinctive qualitative
features of the meson spectrum: mass splitting between pseudoscalar and
vector mesons, Regge trajectories and the asymptotic mass formulas in the
heavy quark limit. Numerically, this model describes to within ten percent
accuracy the masses and weak decay constants of mesons.

A relativistic constituent quark model developed first in \cite{efim88} has found
numerous applications both in the meson sector (e.g., \cite{ivan06}) and in
baryon physics (e.g., \cite{faes09}). In the latter case baryons are considered
as relativistic systems composed of three quarks. The next step in the
development of the model has been done in \cite{bran09}, where infrared
confinement for a quark-antiquark loop was introduced. The implementation of
quark confinement allowed to use the same values for the constituent quark
masses both for the simplest quark-antiquark systems (mesons) and more
complicated multiquark configurations (baryons, tetraquarks, etc.). Recently,
a smooth decreasing behavior of the Fermi coupling on mass scale has been
revealed by considering meson spectrum within this model \cite{ganb15}.

In a series of papers~\cite{ganb02,ganb09,ganb10,ganb12} relativistic models
with specific forms of analytically confined propagators have been developed to
study different aspects of low-energy hadron physics. Particularly, the role of
analytic confinement in the formation of two-particle bound states has been
analyzed within a simple Yukawa model of two interacting scalar fields, the
prototypes of 'quarks' and 'gluons'. The spectra of the' two-quark' and 'two-gluon'
bound states have been defined by using master constraints similar to the ladder
Bethe-Salpeter equations. The 'scalar confinement' model could explain the
asymptotically linear Regge trajectories of 'mesonic' excitations and the
existence of massive 'glueball' states ~\cite{ganb02}. An extension of this model
has been provided by introducing color and spin degrees of freedom, different
masses of constituent quarks and the confinement size parameter that resulted in
a estimation of  the meson mass spectrum (with relative errors $<3.5$ per cent)
in a wide energy range \cite{ganb09}. As a further test, the weak decay constants
of light mesons and the lowest-state glueball mass has been esimated with
reasonable accuracies. Then, a phenomenological model with specific forms of
infrared-confined propagators has been developed to study the mass-scale
dependence of the QCD effective coupling $\alpha_s$ at  large distances
~\cite{ganb10,ganb12}. By fitting the physical masses of intermediate and heavy
mesons we predicted a new behavior of $\alpha_s(M)$ in the low-energy domain,
including  a new, specific and finite behavior of $\alpha_s(M)$ at origin. Note,
$\alpha_s(0)$ depended on $\Lambda$, we fixed $\alpha_s(0)=0.757$ for
$\Lambda=345$ MeV in \cite{ganb10}.

In the present paper, we propose a new insight into the phenomena of strong
running coupling and hadron mass generating by introducing infrared confined
propagators within a QCD-inspired relativistic field model. First, we derive
a meson mass master equation similar to the ladder Bethe-Salpeter equation
and study a specific new behaviour of the mass-dependent strong coupling
$\hat\alpha_s(M)$ in the time-like region. Then, we estimate properties of the
lowest-state glueball, namely its mass and 'radius'. The spectrum of conventional
mesons are estimated by introducing a minimal set of parameters. An accurate
estimation of the leptonic decay constants of pseudoscalar and vector mesons
is also performed.

The paper is organized as follows. After the introduction, in Section II we give
a brief sketch of main structure and specific features of the model, including
theultraviolet regularization of field and strong charge as well as the infrared
regularizations of the propagators in the confinement domain. A self-consistent
mass-dependent effective strong coupling is derived and investigated in
Section III. The formation of an exotic di-gluon bound state, the glueball,
its ground-state properties are considered in Section IV. Hereby we fix the
global parameter of our model, the confinement scale $\Lambda=236$ MeV.
In Sections V and VI  we give the details of the calculations for the the mass
spectrum and leptonic (weak)  decay constants of  the ground-state mesons
in a wide range of scale. Finally, in Section VII we summarize our findings.

\section{MODEL}

Consider the gauge invariant  QCD Lagrangian:
\begin{equation}
{\cal L}=-\frac{1}{4}\left(
\partial^\mu {\cal A}_\nu^A-\partial^\nu {\cal A}_\mu^A
- g f^{ABC}{\cal A}^B_\mu {\cal A}^C_\nu
 \right)^2
+\left( \bar{q}^a_f\left[ \gamma_\alpha \partial^\alpha-m_f \right]^{ab}
q^b_f \right) + g \left( \bar{q}^a_f \left[\Gamma^\alpha_C {\cal A}^C_\alpha
\right]^{ab}q^b_f \right)\,,
\label{lagrangian}
\end{equation}
where ${\cal A}^C_\alpha$ is the gluon field, $q_f^a$ is a quark spinor of flavor
$f$ with color $a=\{1,2,3\}$ and mass $m_f \!\!=\!\! \{m_{ud},m_s,m_c,m_b\}$,
$\Gamma^\alpha_C \!=\! i\gamma_\alpha t^C$ and $g$ - the strong coupling strength.

Below we study two-particle bound state properties within the model.
The leading-order contributions to the spectra of quark-antiquark and di-gluon
bound states are given by the partition functions:
\begin{eqnarray}
&& Z_{q\bar{q}} = \int\!\!\!\int\!\!{\mathcal{D}}\bar{q}{\mathcal{D}} q
\exp\left\{ -(\bar{q} S^{-1}q)+\frac{g^2}{2} \left\langle
(\bar{q}\Gamma{\cal A} q )
(\bar{q}\Gamma{\cal A} q ) \right\rangle_D   \right\} \,, \\
&& Z_{{\cal AA}} = \left\langle \exp\left\{-\frac{g}{2}
\left( f {\cal A} {\cal A} F \right)\right\} \right\rangle_D\,, \qquad
\langle  (\bullet) \rangle_D \doteq \int\!\!{\mathcal{D}} {\cal A}
~e^{-\frac{1}{2}({\cal A} D^{-1}{\cal A})} (\bullet)\,.
\label{pathint}
\end{eqnarray}

Our first step is to transform these partition functions so that they could be
rewritten in terms of meson and glueball fields.

Let's first evaluate quark-antiquark bound states.  By omitting intermediate
calculation details which can be found in \cite{ganb09,ganb10} we
transform $Z_{q\bar{q}}$ into a new path integral written in terms of meson
fields $B_{\cal N}$ as follows:
\begin{eqnarray}
\label{pathmeson}
Z_{q\bar{q}} \rightarrow  Z_B  = \int {\prod\limits_{\cal N}  {\mathcal{D}}
B_{\cal N}  \,\exp  \left\{ { - \frac{1}{2}  \sum\limits_{{\cal NN'}}  (B_{\cal N} \,
 [\delta^{{\cal NN'}} - \alpha_s\lambda_{\cal NN'} ]\,B_{{\cal N'}} )
 + W_{res} [g\,B_{\cal N} ]} \right\}} \,,
\end{eqnarray}
where all quadratic field configurations ($\sim g^2\,B^2_{\cal N})$ are isolated
in the 'kinetic' term mostly defined by the LO kernel of the polarization operator
of meson $\lambda_{\cal NN'}(z,x,y)$ and interaction between mesons are
described by $ W_{res} [g\,B_{\cal N}] \sim 0(g^3\,B_{\cal N}^3)$. Here
$\alpha_s\doteq g^2/4\pi$ and ${\cal N}=\{Q,J,f_1,f_2\}$ with
$Q=\{n,l,\{\mu\}\}$ - a set of radial $n$, orbital $l$ and magnetic
$\{\mu\}=(\mu_1,...,\mu_l)$ quantum numbers.

\subsection{UV Regularization of Meson field and Strong Charge}

It is a difficult problem to describe a composite particle within
QFT which operates with free fields quantized by imposing
commutator relations between creation and annihilation operators.
The asymptotic {\it in-} and {\it out-} states are constructed by means of
these operators acting on the vacuum state. Physical processes are
described by the elements of the S-matrix taken for the relevant
{\it in-} and {\it out-} states. The original Lagrangian requires renormalization,
i.e. the transition from unrenormalized quantities like mass, wave function,
coupling constant to the physical or renormalized ones.

Let us consider a system of orthonormalized basis functions $\{U_Q(x)\}$:
\begin{eqnarray}
\int dx~U_{Q}(x)U_{Q'}(x)=\delta_{QQ'},~~~~~~~
\sum\limits_{Q}U_{Q}(x)U_{Q'}(y) = \delta(x-y) \,,
\end{eqnarray}
 Particularly,
it may read as:
\begin{eqnarray}
\label{orthonormal}
U_{nl\{\mu\}}(x) \sim T_{l\{\mu\}}(x) \,
L_n^{(l+1)}\left(2c x^2\right) \, e^{- c x^2} \,,
\end{eqnarray}
where $c>0$ is a parameter,  $T_{l\{\mu\}}$ is spherical  harmonics and
$L_n^{(l+1)}(x)$ are the Laguerre polynomials.

Then, the  Fourier transform of the  polarization kernel may be diagonalized
on $\{U_Q(x)\}$ as follows:
$$
\int\!\!\!\int\!\! dx dy \, U_Q(x) \lambda_{\cal NN'}(p,x,y) ~U_Q(y)
=\delta^{\cal NN'}~\lambda_{{\cal N}}(-p^2)
$$
that is equivalent to the solution of the corresponding ladder BSE. Here,
\begin{eqnarray}
\label{lambda}
\lambda_{{\cal N}}(-p^2) &=& \frac{8\, C_J}{9\pi^3}
\int \!\! {d^4 k} \left| V_J(k) \right|^2 \Pi_{\cal N}(k,p) \,,
\end{eqnarray}
and a vertex function and the polarization kernel are defined as follows:
\begin{eqnarray}
\label{vertex}
V_J(k) &\doteq& \int\!\! d^4x\, U_J(x) \sqrt{D(x)}\, e^{-ikx} \,, \\
\Pi_{\cal N}(k,p) &\doteq& - \frac{N_c}{4!} \, {\rm Tr}\left[O_J \tilde{S}_{m_1}
\left(\hat{k}+\xi_1\hat{p} \right) O_{J'} \tilde{S}_{m_2}
\left(\hat{k}-\xi_2\hat{p}\right) \right]\,. \nonumber
\end{eqnarray}
Here, $N_c=3$,  the trace is taken on spinor indices, $C_J=\{1,1,1/2,-1/2,0\}$ are
the Fierz  coefficients of the different spin combinations
$O_J=\{I,i\gamma_{5}, i\gamma_{\mu},\gamma_{5} \gamma_{\mu},
i[\gamma_\mu,\gamma_\nu]/2 \}$ for {\sl scalar, pseudoscalar, vector
et cet.} meson states $J=\{S,P,V,A,T\}$.

The gluon $\tilde{D}(p)$ (in Feynman gauge) and quark propagator
$\tilde{S}_{m_1}(\hat{p})$ defined in Euclidean momentum space read:
\begin{eqnarray}
&&
\tilde{D}_{\mu\nu}^{AB}(p)=\delta^{AB} \delta_{\mu\nu} \cdot\tilde{D}_0(p),  \qquad \qquad 
\tilde{D}_0(p) = \frac{1}{p^2}  =  \int\limits_0^{\infty} \!\!\! ds \, e^{-s p^2} \,, \nonumber\\
&&
\tilde{S}_{m_f}(\hat{p}) \!=\! \frac{1}{-i\hat{p}+m_f}
=( i\hat{p}+m_f) \cdot\tilde{S}^0_{m_f}(p), \qquad 
\tilde{S}^0_{m_f}(p) = \!\!\int\limits_{0}^{\infty} dt \exp[-t(p^2+m_f^2)] \,.
\label{propagators}
\end{eqnarray}

In relativistic quantum-field theory a stable bound state of $n$ massive
particles shows up as a pole in the S-matrix with a center of mass energy.
Accordingly, we go into the meson mass shell $-p^2=M_J^2$ and expand
the quadratic term in Eq. (\ref{pathmeson}) as follows:
\begin{eqnarray}
\label{renormed}
(B_{\cal N}[1-\alpha_s\lambda_{\cal N}(-p^2)]B_{\cal N})
=(B_{\cal N}[1-\alpha_s\lambda_{{\cal N}}(M_{{\cal N}}^2)
-\alpha_s\dot\lambda_{\cal N}(M_{\cal N}^2)[p^2+M_{\cal N}^2] B_{\cal N})  \,.
\end{eqnarray}
Then, we rescale the boson field and strong charge as
\begin{eqnarray}
B_{\cal N}(x) \to B_R(x) / \sqrt{\alpha_s \dot\lambda_{\cal N}(M_{\cal N}^2)} \,,
\qquad
g \, B_{\cal N}(x) \to g_R \, B_R(x) \,,
\qquad
\dot\lambda_{\cal N}(z) \doteq \frac{d\lambda_{\cal N}(z)}{dz}
\label{renorm1}
\end{eqnarray}
If we require a condition
\begin{eqnarray}
1-\alpha_s\lambda_{{\cal N}}(M_{{\cal N}}^2)=0
\label{massequa}
\end{eqnarray}
one obtains the Lagrangian of meson field $B_R$ with the mass $M$
and Green's function $\left(p^2+M_{\cal N}^2\right)^{-1}$ in the fully
renormalized partition function (the conventional form) as follows:
\begin{eqnarray}
 Z  = \int \! {\mathcal{D}}B_R  \, \exp  \left\{
 - \frac{1}{2} \left(B_R \left[  p^2+M_{\cal N}^2 \right]  B_R \right)
 + W_{res} [g_R\,B_R ] \right\}  \,.
\end{eqnarray}
It is easy to find that regularizations (\ref{renorm1}) lead to another
requirement:
\begin{eqnarray}
Z_M=1-\alpha_R \dot\lambda_{{\cal N}}(M_{{\cal N}}^2)=0 \,,
\label{composit}
\end{eqnarray}
that is nothing else but the 'compositeness' condition (see, e.g. \cite{jouv56})
which means that the renormalization constant of the mesonic field $Z_M$
is equal to zero and bare meson fields are absent in the consideration.

Since the calculation of the Feynman diagrams proceeds in the Euclidean
region where $k^2=-k^2_E$, the vertex function $V_J(k)$ decreases rapidly
for $k^2_E\to\infty$ and thereby provides ultraviolet convergence in the
evaluation of any diagram.

\subsection{IR Regularization of the Green Functions}

Ultraviolet singularities in the model have been removed by renormalizations
of wave function and charge, but infrared divergences remain in
Eq.(\ref{massequa}) because of propagators in Eq.(\ref{propagators}).
The QCD vacuum structure remains unclear and the definition of the explicit
quark and gluon propagator encounters difficulties in the confinement region.
Particularly, IR behaviors of  the quark and gluon propagators are not
well-established and need to be more specified \cite{shir02}.
It is clear that conventional forms of the propagators in Eq.(\ref{propagators})
cannot adequately describe the hadronization dynamics and the currents and
vertices used to describe the connection of quarks and gluons inside hadrons
cannot be purely local.
Nowadays, any widely accepted and rigorous analytic solutions to these
propagators are still missing.

In our previous papers specific forms of quark and gluon propagators
were exploited \cite{ganb09,ganb10}. These propagators were entire
analytic functions in Euclidean space and represented simple and
reasonable approximations to the explicit propagators calculated in the
background of vacuum gluon field obtained in \cite{efim95}.

On the other hand, there are theoretical results predicting an IR behavior
of the gluon propagator. Particularly, a gluon propagator propagator was
inversely proportional to the dynamical gluon mass \cite{alle96} at the
momentum origin $p^2=0$, while others equaled  to zero \cite{fisc02,lerc02}.
Numerical lattice studies \cite{lang02} and renormalization group analysis
\cite{gies02} also indicated an IR finite behavior of gluon propagator.

Below we follow these theoretical predictions in favor of an IR-finite behavior
of the gluon propagator and exploit a scheme of 'soft' infrared cutoffs on
the limits of scale integrations for the scalar parts of both propagators as follows:
\begin{eqnarray}
&&
D_0 (x)  \!=\! \frac{1}{(2\pi)^2 x^2} \to
D_{\Lambda}(x) = \frac{1}{(2\pi)^2} \int\limits_{0}^{1/\Lambda^2} \!\!ds 
\, \frac{\exp[-x^2/(4s)]}{s^2} \,, \nonumber\\
&&
\tilde{S}^0_{m_f}(p)  \to \tilde{S}^{\Lambda}_{m_f}(p) =
\!\!\int\limits_{0}^{1/\Lambda^2} \!\!dt \, \exp[-t(p^2+m_f^2)] \,.
\label{confined}
\end{eqnarray}
Propagators $D_{\Lambda}(x)$ and $\tilde{S}^{\Lambda}_{m_f}(p)$ do not  
have any singularities in the finite $x^2$- and $p^2$- planes in Euclidean space,
thus indicating the absence of a single gluon (quark) in the asymptotic space
of states. An IR parametrization is hidden in the energy scale $\Lambda$ of 
confinement domain. The analytic confinement disappears as $\Lambda\to 0$. 
Note, propagators in Eq.(\ref{confined}) differ from those used previously in 
\cite{ganb02,ganb09,ganb10,ganb12} and represent lower bounds to the explicit 
ones.

\subsection{Meson Mass Equation}

The dependence of meson mass $M$  on $\alpha_s$ and other model
parameters $\{\Lambda, m_1, m_2\}$ is defined by Eq. (\ref{massequa}).
Further, it is convenient to go to dimensionless co-ordinates, momenta
and masses as follows:
\begin{eqnarray}
\label{relative}
x_\nu\cdot\Lambda \to x_\nu\,, \quad
k_\nu/\Lambda \to k_\nu\,, \quad p_\nu/\Lambda \to p_\nu\,, \quad
m_1/\Lambda \to \mu_1\,, \quad  m_2/\Lambda \to \mu_2\,, \quad
M/\Lambda \to \mu\,, \quad  \nu=\{1,2,3,4\}\,.
\end{eqnarray}

The polarization kernel  $\lambda_{{\cal N}}(-p^2)$ in Eq. (\ref{lambda})
is natively obtained real and symmetric that allows us to find a simple
variational solution to this problem. Choosing a trial Gaussian function
for the ground state mesons:
\begin{eqnarray}
\label{ground}
U(x,a) = \frac{2a}{\pi} \exp\left\{-{a x^2}\right\}\,,
\qquad \!\! \int\!\! d^4x \left| U(x,a)\right|^2 =1\,, \qquad a>0
\end{eqnarray}
we obtain a variational form of equation (\ref{massequa}) for meson
masses as follows:
\begin{eqnarray}
\label{variat}
1 = \alpha_s \cdot \max_{a>0} \lambda (\mu,\mu_1,\mu_2,a)
   = \alpha_s \cdot \hat{\lambda} (\mu,\mu_1,\mu_2) \,.
\end{eqnarray}

Further we exploit Eq. (\ref{variat}) in different ways, by solving either
for $\alpha_s$ at fixed masses $\{\mu,\mu_1,\mu_2\}$,  or
for $\mu$ by keeping $\alpha_s$ and $\{\mu_1,\mu_2\}$ fixed.

\section{EFFECTIVE STRONG COUPLING IN THE IR REGION}

Understanding of both high energy and hadronic phenomena is necessary
to know the strong coupling in the nonperturbative domain at low mass scale
\cite{barn82,higa84,brod02}. Despite important results and constraints
obtained from experiments, most investigations of the IR-behavior of
$\alpha_s$ have been theoretical, a number of approaches have been
explored with their own benefits, justifications and limitations.

The QCD coupling may feature an IR-finite behavior (e.g., in
\cite{agui04, brod04}). Particularly, the averaged IR value of strong coupling
obtained from analyzing jet shape observables in $e^+ e^-$ annihilation is finite
and modest: $\langle\alpha_s\rangle = 0.47\pm 0.07$ for the energy interval
$E< 2$ GeV \cite{doks98}. The stochastic vacuum model approach to high-energy
scattering found that $\alpha_s\sim 0.81$  in the IR region \cite{shos03}. Some
theoretical arguments lead to a nontrivial IR-freezing point, particularly, the
analytical coupling freezes at the value of $4\pi/\beta_0$ within one-loop
approximation \cite{shir97}. The phenomenological evidence for $\alpha_s$ finite
in the IR region is much more numerous \cite{beth00,beth06,cour11,crew15,deur16}.

There is an indication that the most fundamental Green's functions of QCD, such
as the gluon and quark propagators may govern the detailed dynamics of the
strong interaction and the effective strong charge $g$ \cite{agui09}. Therefore,
in the present paper we perform a new investigation of the IR behavior of
$\alpha_s$ as a function of mass scale $M$ by using the IR-confined propagators
defined in Eq.(\ref{confined}) .

In our previous investigation, we studied the mass-scale dependence of
$\alpha_s(M)$ within another realization of analytical confinement and determined
it by fitting physical masses of mesons \cite{ganb10,ganb12}. This strategy
led to a smooth decreasing behaviour of $\alpha_s(M)$, but the result was
depending on a particular choice of model parameters, namely, the masses
$m_1$ and $m_2$ of two constituent quarks composing a meson.

However, any physical observable, including $\alpha_s$, should not depend on
the particular scheme of calculation, by definition. This kind of dependence is
most pronounced in leading-order QCD and often used to test and specify
uncertainties of theoretical calculations for physical observables. There is no
common agreement of how to fix the choice of scheme.

Our idea is to investigate the behaviour of the strong effective coupling
$\alpha_s$ only in dependence of mass scale $\mu$ by solving Eq. (\ref{variat}).
In doing so, the dependencies on  $\mu_1$ and $\mu_2$ may be removed
by revealing and substituting indirect dependencies of $\mu_i=\mu_i(\mu)$.

For this purpose we analyze the meson masses estimated in \cite{ganb10,ganb12}
in dependence of fixed parameters $m_1$ and $m_2$, there. Then, one may easily
notice a pattern: $m_1+m_2 > M$ for light ($\pi$ and $K$) and $m_1+m_2 < M$ for
other mesons ($\rho,~K^*$, ...,~$\eta_b,~\Upsilon$). Hereby, the constituent quark
masses $\{m_{ud}, m_s, m_c, m_b\}$ were obtained by fitting $\alpha_s(M)$ at
physical masses of $\{D^*,~D_s^*,~J/\Psi,~\Upsilon\}$ and then, we calculate
$(m_1+m_2)/M=\{0.816,~0.824,~0.935,~0.992\}$ of these mesons, consequently.

A similar pattern is also revealed in the case of our earlier model with 'frozen'
strong coupling not depending on mass scale \cite{ganb09}.  Also, it was stressed
that the self-energy function $\lambda(M,~m_1,~m_2)$ was low sensitive under
significant changes of parameters $m_1,~m_2$  (see Fig.2 in \cite{ganb09}).

Therefore, not loosing the general pattern, we can substitute an 'average'
dependence $m_1=m_2=M/2$. As mentioned above, this assumption is not able
to change drastically the behaviour of $\alpha_s$. Controversaly, we now define
$\alpha_s$ more self-consistently, in dependence only on the mass scale
$\mu=M/ \Lambda$ by eliminating the direct presence of constituent quark masses.

Finally, we calculate a variational solution $\hat{\alpha}_s$  to $\alpha_s$ in
dependence on a dimensionless energy-scale ratio $\mu$ as follows:
\begin{eqnarray}
\label{running}
\hat{\alpha}_s(\mu)= 1/ \hat{\lambda} (\mu,\mu/2,\mu/2) \,.
\end{eqnarray}

\begin{figure}[h]
\includegraphics[width=70mm,height=55mm]{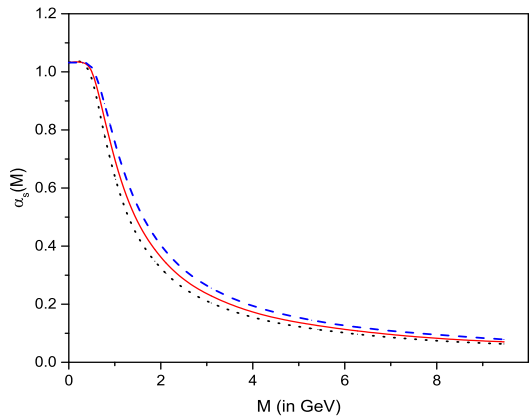} \hspace{1cm}%
\begin{minipage}[b]{70mm}
\caption{\label{Fig1}
Mass-dependent effective strong coupling $\hat{\alpha}_s(M)$
for different values of confinement scale (dots for $\Lambda$=216 MeV,
solid line for $\Lambda$=236 MeV and dashes for $\Lambda$=256 MeV}
\end{minipage}
\end{figure}

The behaviour of new variational upper bound $\hat{\alpha}_s(\mu)$ to
$\alpha_s(\mu)$ is plotted in Fig.1. The slope of the curve depends on
$\Lambda>0$, but the value at origin remains fixed for any $\Lambda>0$
and equals to
\begin{eqnarray}
\hat{\alpha}_s(0) = \hat{\alpha}_s^0=1.03198 \,,
\qquad \mbox{or} \qquad {\hat{\alpha}_s^0/\pi}
=0.328489\,.
\label{running0}
\end{eqnarray}

We use the meson mass $M$ as the appropriate characteristic parameter, so the
coupling $\hat{\alpha}_s(M)$ is defined in a time-like domain ($s=M^2$). On the
other hand, the most of known data on $\alpha_s(Q)$ is possible in space-like
region \cite{PDG16}. The continuation of the invariant charge from the time-like
to the  spacelike region (and vice versa) is elaborated by making use of the
integral relationships (see, e.g. \cite{milt97}). Particularly, there takes place
a relation \cite{nest03}:
\begin{eqnarray}
\alpha_s(q^2)=q^2\int\limits_{0}^{\infty} \frac{ds}{(s+q^2)^2} \, \hat{\alpha}_s(s)
\label{spacetimelike}
\end{eqnarray}
A detailed study of this transformation deserves a separate consideration and
below we just note that at origin ($q^2=-s=0$) both representations converge:
\begin{eqnarray}
\alpha_s(0)=\hat{\alpha}_s(0) \int\limits_{0}^{\infty} \frac{dt}{(1+t)^2}
= \hat{\alpha}_s(0)\cdot 1\,.
\end{eqnarray}

Therefore, the freezing value $\hat{\alpha}_s^0=1.03198$ may be compared
with those obtained as continuation of $\alpha_s(Q)$ in space-like domain.
Particularly, in the region below the $\tau$-lepton mass the strong coupling
value is expected between $\alpha_s(M_\tau) \approx 0.34$ \cite{PDG14}
and an IR fix point $\alpha_s(0)=2.972$ \cite{fisch05}. Moreover, a use of
$\overline{MS}$ renormalization scheme leads to value
$\alpha_s(0)=1.22 \pm 0.04 \pm 0.11 \pm 0.09$ for confinement scale
$\Lambda_{QCD}=0.34 \pm 0.02$ GeV \cite{deur16}.

We conclude that our IR freezing value $\hat{\alpha}_s(0)=1.03198$ is in
a reasonable agreement with above mentioned predictions and does not
contradict other quoted estimates:
\begin{equation}
\left\{
\begin{tabular}{l}
${\alpha_s^0/\pi} \simeq 0.19 - 0.25$ \qquad  \cite{godf85}\,, \\
${\alpha_s^0/\pi}\simeq 0.265$ \qquad \qquad ~~\cite{zhan91}\,, \\
${\alpha_s^0/\pi}  \simeq 0.26$ \qquad \qquad ~~~\cite{halz93}\,, \\
$\left\langle  {\alpha_s^0/\pi}  \right\rangle_{1\,GeV}  \simeq 0.2$
\qquad ~~~\cite{doks96}
\end{tabular}
\right.
\label{alpha0IR}
\end{equation}
and phenomenological evidences \cite{bald02,bald08}.

It is important to stress that we do not aim to obtain the behavior of the
coupling constant at all scales. At moderate $M^2=-p^2$ we obtain $\alpha_s$
in coincidence with the QCD predictions. However, at large mass scale
(above 10 GeV) $\hat{\alpha}_s$  decreases faster. The reason is the use
of confined propagators in the form of entire functions in Eqs.~(\ref{confined}).
Then, the convolution of entire functions leads to a rapid decreasing in Euclidean
(or,  a rapid growth in Minkowskian) space of physical matrix elements once the
mass and energy of the reaction have been fixed. Consequently, the numerical
results become sensitive to changes of model parameters at large masses and
energies.

Note, any physical observable must be independent of the particular scheme
and mass by definition, but in (\ref{running}) we obtain  $\hat\alpha_s$ in
dependence on scaled mass $M/\Lambda$. This kind of scale dependence
is most pronounced in leading-order QCD and often used to test and specify
uncertainties of theoretical calculations for physical observables.
Conventionally, the central value of $\alpha_s(\mu)$ is determined or taken for
$\mu$ equalling the typical energy of the underlying scattering reaction.
There is no common agreement of how to fix the choice of scales.

Below,  we will fix the model parameter $\Lambda$ by fitting the scalar glueball
(two-gluon bound state) mass.

\section{LOWEST GLUEBALL STATE}

Most known experimental signatures for glueballs are an enhanced production in
gluon-rich channels of radiative decays and some decay branching fractions
incompatible with $(q\bar{q})$ states. Particularly, there are predictions expecting
non-$q\bar{q}$ scalar objects, like glueballs in the mass range  $\sim 1.5\div 1.8$
GeV \cite{amsl04,bugg04,yao06}. Some references favor the $f_0(1710)$ and
$f_0(1810)$ as the lightest glueballs \cite{chan07,abli06}, while heavy  glueball-like
states (pseudoscalar, tensor, ...) are expected in the mass range
$M_G\sim 2.4\div 4.9$ GeV with different spins $J=0, 1, 2, 3$ \cite{PDG16}.

Gluodynamics has been extensively investigated within quenched lattice
QCD simulations. A use of fine isotropic lattices  resulted in a value
~1.475 GeV for the scalar glueball mass \cite{meye05}.  An improved
quenched lattice calculation at the infinite volume and continuum limits
estimates the scalar glueball mass equal to  $1710\pm50\pm80$ MeV
\cite{chen06}.

Among different glueball models, the two-gluon bound states are the most
studied purely gluonic systems in the literature, because when the
spin-orbital interaction is ignored ($\ell =0$), only scalar and tensor states
are allowed. Particularly,  the lightest glueballs with positive charge parity
can be successfully modeled by a two-gluon system in which the constituent
gluons are massless helicity-one particles \cite{math08}.

Below we consider a pure two-gluon scalar bound state with $J^{PC}=0^{++}$.
By omitting details of intermediate calculations (similar to those represented in
the previous section) we define the scalar glueball mass $M_{0^{++}}$ from
equation:
\begin{eqnarray}
1- \frac{8\,\hat{\alpha}}{3\pi} \max_{a>0} \int\! dz \, e^{izp} ~\Pi_{G}(z)=0 \,,
\qquad p^2=-M_{0^{++}}^2 \,,
\label{gluemass}
\end{eqnarray}
where
\begin{eqnarray*}
 \Pi_G(z)\doteq \int\!\!\!\int\!\! dt ds ~U(t,a)  \sqrt{W_{\Lambda}(t)}
~D_{\Lambda}\left(\frac{t+s}{2}+z\right)D_{\Lambda} \left(\frac{t+s}{2}-z\right)
\sqrt{W_{\Lambda}(s)} ~U(s,a)
\end{eqnarray*}
is the self-energy (polarization) function of the scalar glueball and 
$W_{\Lambda}(s)$ is a potential function connecting scalar gluon currents.
The ground state basis $U(t,a)$ may be chosen as in Eq.(\ref{ground}).
Then, we are able to estimate an upper bound to the scalar glueball mass
by using the effective mass-dependent coupling defined in Eq.(\ref{running}).

Our model has a minimal set of free parameters:
$\{\hat{\alpha}, \Lambda,m_{ud},m_s,m_c,m_b\}$. The glueball mass depends
on $\{\hat{\alpha}, \Lambda\}$. We fix $\Lambda$ by fitting the expected
glueball mass. Particularly, for $\Lambda=236$ MeV and $\hat\alpha(M_G)$
defined in Eq.(\ref{running}) we obtain new estimates:
\begin{eqnarray}
M_{0^{++}} = 1739{\mbox{\rm ~MeV}} \,, \qquad \hat\alpha(M_{0^{++}})=0.451 \,.
\label{Glueballmass}
\end{eqnarray}

The new value of $M_{0^{++}}$ in  (\ref{Glueballmass}) is in agreement not
only with our previous estimate \cite{ganb09}, but also with other predictions
expecting the lightest glueball located in the scalar channel in the mass range
$\sim 1500\div 1800{\rm ~MeV}$ \cite{nari00,bali01,amsl04,meye05,greg12}.
The often referred quenched  QCD calculations predict
$1750 \pm 50 \pm 80{\mbox{\rm ~MeV}}$  for the mass of the lightest
glueball \cite{morn99}. The recent quenched lattice estimate with improved
lattice spacing favors a  scalar glueball mass
$M_G=1710\pm 50 \pm 58 {\rm ~MeV}$ \cite{chen06}.

Another important property of the scalar glueball is its size, the 'radius'
which should depend somehow on the glueball mass.  We estimate the
glueball radius roughly as follows:
\begin{eqnarray}
r_{0^{++}} \sim \frac{1}{2\Lambda}
\sqrt{ \frac{\int\! d^4 x~x^2~W_{\Lambda}(x)~U^2(x)}
{\int\! d^4 x ~W_{\Lambda}(x)~U^2(x)}}
\approx \frac{1}{394.3~\mbox{\rm MeV}} \approx  0.51~\mbox{\rm fm} \,.
\label{radius}
\end{eqnarray}

This may indicate that  the dominant forces binding gluons are provided
by vacuum fluctuations of correlation length $\sim 0.5\,\mbox{\rm fm}$.
On the other side, typical energy-momentum transfers inside a scalar
glueball should occur in the confinement domain
$\sim 236\,\mbox{\rm MeV}\sim 0.85\,\mbox{\rm fm}$, rather than
at the chiral symmetry breaking scale
$\Lambda_\chi \sim 1\,\mbox{\rm GeV} \sim 5\,\mbox{\rm fm}$.

From (\ref{Glueballmass}) and (\ref{radius}) we deduce that
$$
r_{0^{++}} \cdot M_{0^{++}} \approx 4.41 \,.
$$
This value may be compared with the prediction
($r_{G} \cdot M_G = 4.16\pm 0.15$) of quenched QCD calculations
\cite{morn99,chen06}.

A quenched lattice QCD study of the glueball properties at finite
temperature with the anisotropic lattice imposes restriction on the
glueball radius at zero temperature:
$0.37\, {\rm fm} < r_G < 0.57\, {\rm fm}$ \cite{iish01} that is in
agreement with our result.

The gluon condensate is a non-perturbative property of the QCD vacuum and
may be partly responsible for giving masses to certain hadrons. The correlation
function in QCD dictates the value of  corresponding condensate. Particularly,
with $\Lambda=236$ Mev and $\hat\alpha_s=0.451$ we calculate the lowest
non-vanishing gluon condensate in the leading-order (ladder) approximation:
$$
\frac{\hat\alpha_s}{\pi} \left\langle F_{\mu\nu}^A  F^{\mu\nu}_A  \right\rangle
=\frac{16 N_c}{\pi} \Lambda^4  \approx 0.0214~ GeV^4
$$
which is in accordance with a refereed value \cite{nari12}
$$
\alpha_s \left\langle G^2  \right\rangle = (7.0\pm 1.3)\cdot 10^{-2} ~ GeV^4 \,.
$$

\section{MESON SPECTRUM}

Below we consider the most established sector of hadron physics, the
spectrum of conventional (pseudoscalar ${\mathbf{P}} (0^{-+})$ and
vector ${\mathbf{V}} (1^{--})$) mesons.

In previous investigations with analytic confinement \cite{ganb09,ganb10,ganb12},
we fixed all the model parameters ($\Lambda, \hat\alpha_s, m_{ud}, m_s, m_c, m_b$)
by fitting the real meson masses.

In the present paper, the universal confinement scale $\Lambda=236 \, MeV$
is fixed by fitting the scalar glueball mass. And the effective strong coupling
$\hat\alpha_s$ is unambiguously determined by Eq.(\ref{running}).

Therefore,  we derive meson mass formula Eq.(\ref{massequa}) by fitting the
meson physical masses with adjustable parameters $\{ m_{ud}, m_s, m_c, m_b\}$.

This results in a new final set of model parameters (in units of MeV)  as follows:
\begin{eqnarray}
\label{fixedparameters}
\Lambda=236\,, ~~m_{ud}=227.6\,, ~~m_s=420.1\,, ~~m_c=1521.6\,,
~~m_b=4757.2 \,.
\end{eqnarray}

\vskip -3mm
\begin{table}[h]
\caption{\label{Tab1}
Estimated masses of conventional mesons $M_P$ and  $M_V$  (in units of
{\rm MeV}) for model parameters (\ref{fixedparameters}) compared to
the recent experimental data \cite{PDG16}.}
\begin{center}
\begin{tabular}{llll}
\hline\noalign{\smallskip}
$0^{-+}~~$  & $M_P~(MeV)~$  &  Data~(MeV)~ \\
\hline
$D$            & 1893.6  & 1869.62   \\
$D_s$        & 2003.7  & 1968.50   \\
$\eta_c$    &  3032.5  &  2983.70  \\
$B$            &  5215.2  & 5259.26   \\
$B_s$        &  5323.6  & 5366.77  \\
$B_c$        &  6297.0  & 6274.5    \\
$\eta_b$    &  9512.5  & 9398.0    \\
\hline\noalign{\smallskip}
$1^{--}~~$  & $M_V~(MeV)~$  &  Data~(MeV)~ \\
\hline
$\rho$        &   774.3   &  775.26    \\
$K^*$         &   892.9   &  891.66    \\
$\Phi$        &  1010.3  &  1019.45  \\
$D^*$         &  2003.8  &  2010.29  \\
$D_s^*$     &  2084.1  &  2112.3    \\
$J/\Psi$     &  3077.6  &  3096.92  \\
$B^*$         &  5261.5  &  5325.2    \\
$B_s^*$     &  5370.9  &  5415.8    \\
$\Upsilon$  &  9526.4  &  9460.30  \\
\noalign{\smallskip}\hline
\end{tabular}
\end{center}
\vskip -5mm
\end{table}

The constituent quark mass values fall into the expected range. The present
numerical least-squares fit for meson masses and the values for the model
parameters supersede our previous results in \cite{ganb09,ganb10} obtained
by exploiting different types of analytic confinement and running coupling.

Note, we consider $\omega$ and $\Phi$ as 'pure' states without mixing.
Also, we pass the $\eta - \eta'$ mixing, because this problem obviously
deserves a separate and complicated consideration due to a possible
gluon admixture to the conventional $q\bar q$-structure of the $\eta'$.

Our present model has only five free parameters ($\Lambda$ and four masses
of constituent quarks) and a constraint self-consistent equation for $\alpha_s$.
Nevertheless, our estimates on the conventional meson masses represented
in  TAB. I are in  reasonable agreement with experimental data and the relative
errors do not exceed $1.8$ per cent in the whole range of mass scale.

\section{LEPTONIC DECAY CONSTANTS OF MESONS}

One of the important quantities in the hadron physics is the leptonic (weak)
decay constant of meson. The precise knowledge of its value provides
significant improvement in our understanding of various processes convolving
meson decays. Particularly, the weak decay constants of light mesons are well
established data and many collaboration groups have these with sufficient
accuracy \cite{bern07,bean07,foll07},

Therefore, the leptonic decay constant values (plotted in Fig.2 in dependence
of meson physical mass) are often used to test various theoretical models.

A given meson in our model is characterized by its mass $M$, two of constituent
quark masses $m_1$  and $m_2$ along the infrared confinement parameter
$\Lambda$ universal for all hadrons, including exotic glueballs. The masses
($m_{ud}$, $m_s$, $m_c$, $m_b$)  of four constituent quarks have been
obtained by fitting the meson physical masses. Hereby, the effective strong
coupling $\hat\alpha_s$ depends on the ratio $M/\Lambda$.

The leptonic decay constants which are known either from experiment
or from lattice simulations is an additional characteristic of a given meson.

We define the leptonic decay constants of pseudoscalar and vector mesons
as follows:
\begin{eqnarray}
i p_{\mu} f_P \!\!\!&=&\!\!\! g_{ren} \frac{8N_c}{9}\int\!\!\! \frac{dk}{(2\pi)^4}
V_P(k) {\rm Tr} \left[ \gamma_\mu (1-\gamma_5) \tilde{S}\left(\hat{k}+\xi_1\hat{p}\right) \,
i \gamma_5 \tilde{S}\left(\hat{k}-\xi_2\hat{p}\right) \right]   \,, \nonumber\\
\delta^{\mu\nu} M_V f_V \!\!\!&=&\!\!\! g_{ren} \frac{8N_c}{9}\int\!\!\! \frac{dk}{(2\pi)^4}
V_V(k) {\rm Tr} \left[ \gamma_\mu \tilde{S}\left(\hat{k}+\xi_1\hat{p}\right) \,
i \gamma_\nu \tilde{S}\left(\hat{k}-\xi_2\hat{p}\right) \right]   \,,
\label{leptonic}
\end{eqnarray}
where $g_{ren}=g/\sqrt{\alpha\dot\lambda(M_J)}$ is the renormalized strong
charge and vertices $V_J(k)$ are defined in Eq.(\ref{vertex}).

\begin{figure}[h]
\includegraphics[width=60mm,height=45mm]{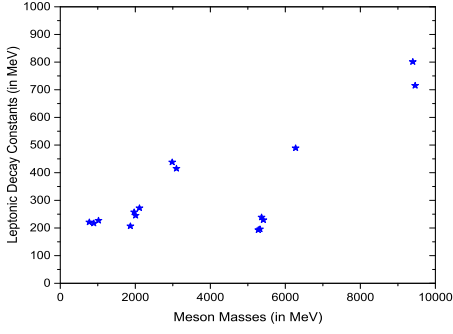} \hspace{1cm}%
\begin{minipage}[b]{70mm}
\caption{\label{Fig2}
Experimental data on leptonic decay constants plotted versus
physical masses of mesons.}
\end{minipage}
\end{figure}


The parameters $\Lambda, M_J, m_1, m_2$ and $\hat\alpha_s$ have
been already fixed by considering the glueball and meson spectra, so
these values in Eq.(\ref{fixedparameters}) will be used to solve
Eqs.(\ref{leptonic}) for $f_P$ and $f_V$.

In doing so we note a 'sawtooth'-type dependence of $f_J$ on meson
masses (see Fig.2) that requires an additional parameterization to model
more adequately this unsmooth behaviour.

For the meson mass equation Eq.(\ref{variat}), the parameter $a$ in the
basis function $U(x,a)$ served as a variational parameter to maximize
the meson self-energy function $\lambda(M,m_1,m_2)$.

In contrast to this, for  Eqs.(\ref{leptonic}) we introduce:
\begin{eqnarray}
\label{ground2}
U(x,R_M) = \frac{R_M^2}{\pi\Lambda^2} \exp\left\{-{R_M^2 x^2}/2\right\}\,,
 \qquad R_M>0\,,
\end{eqnarray}
where $R_M$  characterizes the 'size' of each meson $M$ in units of mass.

Then, we define the meson 'sizes' $R_M$ by solving Eqs.(\ref{leptonic})
with Eq.(\ref{ground2}) and fixed model parameters Eq.(\ref{fixedparameters}).

\begin{table}[h]
\caption{\label{Tab2}
Meson 'size' parameters $R_M$ and leptonic decay constants $f_P$ and $f_V$
compared to experimental data in \cite{PDG14,laih10,chiu07,beci99}.}
\begin{center}
\begin{tabular}{llllll}
\hline\noalign{\smallskip}
$0^{-+}$        & $R_M~(GeV)~$ & $f_P~(MeV)~$  &  Data~(MeV)~ & Ref.                \\
\hline
${D}$         & 0.93   & 207     & 206.7 $\pm$ 8.9  & \cite{PDG14}   \\
${D_s}$     & 1.08   & 257     & 257.5 $\pm$ 6.1  & \cite{PDG14}   \\
${\eta_c}$ & 1.83    & 238     & 238 $\pm$ 8        & \cite{chiu07}    \\
${B}$         & 1.73    & 193     & 192.8 $\pm$ 9.9  & \cite{laih10}    \\
${B_s}$     & 2.18    & 239     & 238.8 $\pm$ 9.5  & \cite{laih10}    \\
${B_c}$     & 3.34    & 488     & 489 $\pm$ 5        & \cite{chiu07}   \\
${\eta_b}$  & 3.80   & 800     & 801 $\pm$ 9        & \cite{chiu07}   \\
\hline
$1^{--}$        & $R_M~(GeV)~$ & $f_P~(MeV)~$  &  Data~(MeV)~ & Ref.                \\
\hline
${\rho}$     & 0.33 & 221  & 221 $\pm$ 1   & \cite{PDG14} \\
${K^*}$      & 0.38 & 217  & 217 $\pm$ 7   & \cite{PDG14} \\
${\Phi}$     & 0.42 & 227  & 227 $\pm$ 2   & \cite{PDG14}  \\
${D^*}$      & 0.78 & 245  & 245 $\pm$ 20 & \cite{beci99}  \\
${D_s^*}$  & 0.90 & 271  & 272 $\pm$ 26 & \cite{beci99}   \\
${J/\Psi}$  & 2.40 & 416  & 415 $\pm$ 7   & \cite{PDG14} \\
${B^*}$      & 3.34 & 196   & 196 $\pm$ 44 & \cite{beci99}   \\
${B_s^*}$  & 0.92 & 228   & 229 $\pm$ 46 & \cite{beci99}  \\
${\Upsilon}$&2.80 & 715   & 715 $\pm$ 5   & \cite{PDG14} \\
\noalign{\smallskip}\hline
\end{tabular}
\end{center}
\end{table}

Note, the 'size' parameters $R_M$ show the expected general pattern: the
'geometrical size' of a meson, which is proportional to $1/R_M$, shrinks
when the meson mass increases.

The obtained values of meson 'sizes' and the best fit values estimated for
the leptonic decay constants are represented in TAB. II.

\section{CONCLUSION}

In conclusion, we demonstrate that many properties of the low-energy phenomena
such as strong running coupling, hadronization processes, mass generation for
quark-antiquark and di-gluon bound states may be explained reasonably within a
QCD-inspired model with infrared confined propagators. We derived a meson mass
equation and by exploiting it revealed a specific new behaviour of the strong coupling
$\hat\alpha_s(M)$ in dependence of mass scale. An infrared-freezing point
$\hat\alpha_s(0)=1.03198$ at origin $M=0$ has been found and it did not depend
on the particular choice of the confinement scale $\Lambda>0$.  A new estimate
of the lowest (scalar) glueball mass has been performed and it was found at
$\sim 1739$ MeV. The scalar glueball 'size' has also been calculated:
$r_G\sim 0.51$ fm. A nontrivial value of the gluon condensate has also been
obtained. We have estimated the spectrum of conventional mesons by introducing
a minimal set of parameters: four masses of constituent quarks $\{u=d,s,c,b\}$ and
$\Lambda$. The obtained values fit  the latest experimental data with relative errors
less than 1.8 percent.  Accurate estimates of the leptonic decay constants of
pseudoscalar and vector mesons have also been performed.

Note, the suggested model in its simple form is far from real QCD. However, our
guess about the structure of the quark-gluon interaction in the confinement region,
implemented by means of confined propagators and nonlocal vertices has been
probed and the obtained numerical results were in reasonable agreement with
experimental data in different sectors of low-energy particle physics. Since the
model is probed and the parameters are fixed, the consideration may be extended
to actual problems in hadron physics, such as spectra of other mesons (scalar,
iso-scalar), higher glueball states, exotic states ($q\bar{q}+gg$ admixtures,
tetraquark, X(3872) and Z(4430), ...), baryon decays
($\Lambda_b \to \Lambda^* + J/\Psi$) etc.

\vskip 1mm
The author thanks S. B.~Gerasimov,  J.~Franklin, M. A.~Ivanov and
Guy F.~de~Teramond for valuable comments and remarks.


\end{document}